%% file: vanlangevelde.tex
\documentclass{evn2004}
\usepackage{txfonts}
\usepackage{graphicx}
\begin{document}
   \title{High output data rates with PCInt on the EVN MkIV data
     processor}

   \author{Huib Jan van Langevelde\inst{1}
          \and
          Harro Verkouter\inst{1}
          \and
          Steve Parsley\inst{1}
          \and
          Mike Garrett\inst{1}
          \and
          Friso Olnon\inst{1}
          \and
          Bauke Kramer\inst{1}
          }

        \institute{Joint Institute for VLBI in Europe, Postbus 2, 7990 AA
          Dwingeloo, the Netherlands
          }

   \abstract{
     The EVN MkIV data processor is being equipped with a high data
     output capability, called PCInt. In its next phase it will allow
     very high data output rates, up to 160 Mb/s. This functionality is
     required for the high spectral resolution that will be offered
     with recirculation. Moreover, VLBI observations traditionally
     focus on a small patch of sky and image typically a few 100 mas
     around a bright source. High spectral and time resolution is
     needed to image a larger area of the sky, up to the primary beam
     of the individual telescopes. After the upgrade such high
     resolution data will become the standard product. From the archive
     of high resolution data it will be possible to image many sources
     in each field of view around the original targets.  
     }

   \maketitle
%

\section{Introduction}

The output data rate of a VLBI processor determines how accurately the
correlation product can be sampled in frequency and time.  The
frequency resolution is important for various spectral line
applications, but it also limits the field of view (FoV) that can be
imaged before bandwidth smearing sets in. The spectral resolution of a
VLBI correlator is usually determined by the computing power built into
its hardware, e.g.\ the number of available lags per baseline.
Furthermore, the product over all available telescope pairs of spectral
resolution and short visibility integration time combines into the
total output rate. There is generally a maximum data rate, determined
by how fast the data can be flushed to a standard computing environment
and saved to disk. This limits not only the spectral sampling but also
the temporal sampling, constraining the FoV further through time
smearing, which scales with baseline length and distance from the field
centre.

Both the spectral capabilities and the output bandwidth of the EVN data
processor at JIVE are being upgraded.  The first by introducing
recirculation and the latter by the PCInt project. In this paper we
discuss the scientific motivation and the future data handling of this
system. We consider the possibility to operate the EVN correlator in
wide field mode for all experiments in order to build up an archive
with large sky coverage, containing a fantastic number of faint
sources.

\section{Scientific Justification} 

A VLBI dataset, if properly calibrated, has flat phase response, both
in time and frequency, for the target position. Positional offsets from
the phase centre introduce increasingly steep phase slopes in frequency
and time. As long as these are properly sampled, sources at the primary
beam edge can still be imaged. Both effects scale with baseline length
(and are therefore particularly severe for VLBI); time smearing also
scales with frequency (Wrobel \cite{wrobel}). The resulting limits on
the FoV for typical VLBI experiments can be seen in Table~\ref{fov}.
Although the original recordings of a single VLBI experiment hold
information over the whole field, typically only $10^3$ out of $10^8$
beams are imaged.

\begin{table}
  \caption[]{The resulting data volume and field of view 
   (by time and bandwidth smearing)
   for a rather modest VLBI recording at 18cm on 8 EVN antennas with
   a total bandwidth of 64 MHz (2 bit sampled). The primary beam of a 25m
   telescope measures 27'. The requirements become even more severe at
   higher spatial resolution (global baselines, or higher frequency).
}
  \label{fov}
  $$
  \begin{array}{p{0.25\linewidth}cccccc}
    \hline
    \noalign{\smallskip}
    Application & N_{\rm sp} & t_{\rm int}  & Output & 
  FoV_{\rm t} & FoV_{\rm bw} & V_{\rm 12hr} \\
  &  & [s] & [MB/s] & ['] & ['] & [GB] \\
  \noalign{\smallskip}
  \hline
  \noalign{\smallskip}
  Traditional      &   128  & 4.000 & 0.02  & 0.70  & 0.82 &   1 \\
  Operational max  &  1024  & 0.500 & 1.50  & 5.57  & 6.59 &  63 \\
  Phase 0          &  2048  & 0.250 & 6.00  &11.14  &13.19 & 253 \\
  Full system      &  4096  & 0.031 &96.00  &89.11  &26.36 &4050 \\
  \noalign{\smallskip}
  \hline
\end{array}
$$
\end{table}

\begin{figure*}
\label{methanol}
\centering
\includegraphics[angle=0,width=0.78\linewidth]{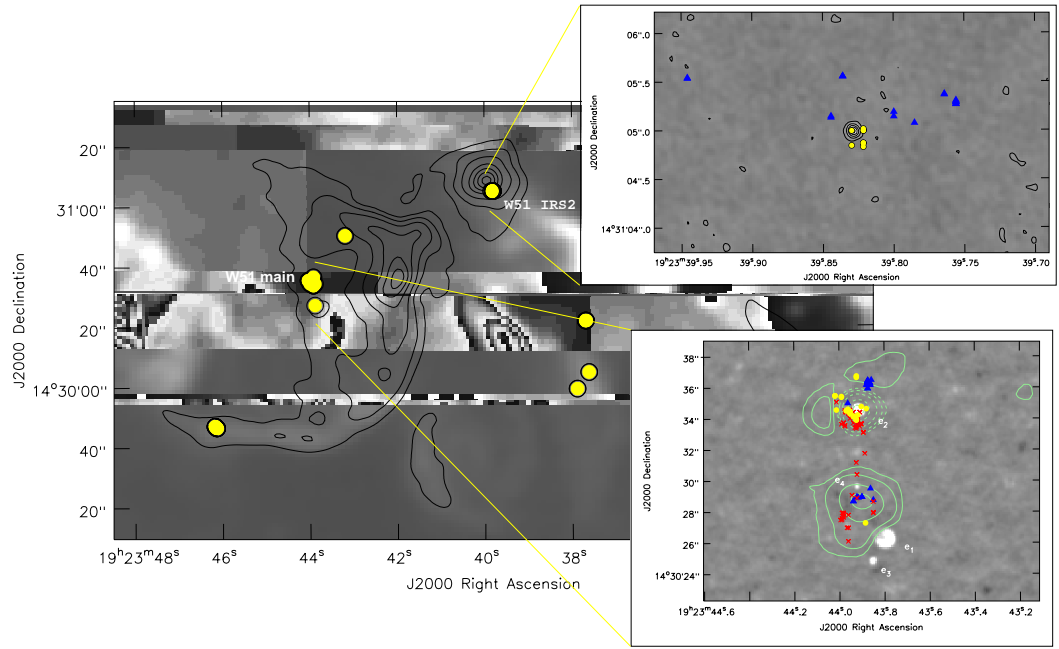}
\caption{The methanol masers in the massive star formation complex
  W51 span arc-minutes on the sky. Imaging these accurately over the
  primary beam of the contributing telescopes requires high spectral
  and temporal resolution. The high resolution data can be compared
  with other tracers of star formation with milli-arcsecond 
  accuracy (Phillips \& van Langevelde \cite{phillips}).
}
\end{figure*}

There are several astronomical applications for which larger fields
of view are useful. Galactic masers may extend over rather large areas, especially
in star formation complexes (Fig.~\ref{methanol}). Gravitational lenses
are another case where VLBI sources may extend over a large FoV.

Moreover, studies of the faint radio source population may be done more
efficiently when a large instantaneous field of view is available.  A
long integration at the full recording bandwidth can then be employed
to study many weak sources simultaneously.  This technique was explored
with the EVN at 1.6 GHz and the VLBA correlator by Garrett et al.
(\cite{garrett}). Even at the moderate resolution of EVN only
observations at 18cm, the FoV is a fraction of the Effelsberg beam, and
barely covers the Hubble Deep field.  Studies like this will benefit
greatly from the upgrades ongoing in the EVN, both in recording and
correlator capacity.

\section{Correlator data flow} 

The EVN MkIV data processor correlates inputs from 16 stations
simultaneously (Schilizzi et al.\ \cite{schilizzi}). Each telescope
input can handle up to 1 Gbit/s, from Mk4 tape, Mk5 disk playback or
fiber connections. Its computing power is based on 32 boards, each
equipped with 32 custom made chips, each producing 256 complex lags.
This yields sufficient spectral capacity to attribute 512 spectral
channels to every baseline between 16 telescopes. With the introduction
of recirculation the number of spectral channels scales up when the
bandwidth of individual sub-bands is below the maximum of 16 MHz.

In its original configuration the data was flushed out on 4 parallel
10Mb/s Ethernet lines.  The first improvement to this system has been
implemented by upgrading the system with 8 Single Board Computers
(SBC), which handle the data-streams and are outfitted with 100 Mb/s
Ethernet to flush the data (Phase 0). In the final PCInt configuration
each rack will have two Single Board Computers running Linux, which
read the data from DSP powered serial ports. A total of $8\times 1$
Gb/s Ethernet connections are then available. The software is modified
in order to process the output data in parallel by a cluster of
workstations.  Special hardware is procured to provide the required IO
bandwidth to disk. In table~\ref{fov} the Field of View (FoV) limits
set by the bandwidth sampling (bw) and time smearing (t) in various
stages of the project are shown.

This system will also be crucial for future spectral line studies.
Currently, spectral resolution often has to be traded for the number of
telescopes, polarization products and time resolution, resulting
sometimes in multiple-pass correlation for large spectral line
projects. With recirculation the spectral capacity could increase by
a factor 8 for some projects (for example OH lines that can be observed
with 2 MHz bandwidth). PCInt will be indispensable for handling such
projects.

\section{User products} 

With the enhanced data rates it is anticipated that all correlation
will one day run at full resolution in order to build up an archive
which covers a substantial area of the sky. Already considerable effort
has gone into making the archive of user products and diagnostic plots
accessible through a web interface: {\tt http://archive.jive.nl/}.  The
data is public after the original proposers grace period. In the future
one can imagine that if the archive contains large FoV data, it may be
interesting for studying e.g.\ the faint source population.  While the
output data will be at full resolution, the user may prefer the data at
a coarser resolution, one that is optimal for the scientific goal of
his study.  An interface to the archive will be designed in order to
allow users to make a selection and create a dataset at a lower
resolution, possibly by averaging for a new target position.  The goal
is to have sufficient (parallel) computing infrastructure, in order to
make such products available in an almost interactive manner.  It is
thought that wide field of view VLBI images will be made as an integral
part of the data processor product.  Such a solution would fit in with
the Virtual Observatory paradigm. Some of the effort for this,
including the streamlining of calibration, is part of the RadioNet
software project ALBUS.

\begin{acknowledgements}
The European VLBI Network is a joint facility of European, Chinese, 
South African and other radio astronomy institutes funded by their 
national research councils.
This research was supported by the European Commission's I3 Programme
``RADIONET", under contract No.\ 505818.
\end{acknowledgements}

\end{document}